# SONOLUMINESCENCE: NATURE'S SMALLEST BLACKBODY

G. Vazquez, C. Camara, S. Putterman, K. Weninger

*Physics Department, University of California, Los Angeles, CA 90095*

The Spectrum of the light emitted by a sonoluminescing bubble is extremely well fit by the spectrum of a blackbody. Furthermore the radius of emission can be smaller than the wavelength of the light. Consequences, for theories of sonoluminescence are discussed.

The phenomenon of sonoluminescence occurs when a trapped bubble of gas in water is driven into high amplitude pulsations by a strong sound wave. Key parameters that characterize the bubble and sound field are the amplitude of the sound $P_a$ and the ambient radius of the bubble $R_0$. This is the radius when $P_a = 0$. During the rarefaction portion of the cycle of sound the bubble expands from $R_0$ to $R_m$ the maximum radius that is about $10R_0$. During the ensuing compression the bubble catastrophically collapses right through $R_0$ on its way down to its collapsed radius $R_c$ that is determined by the van der Waals hard core of the particular gas molecule [for Helium $R_c = R_0/10$]. As R approaches $R_c$ the input acoustic energy density concentrates by 12 orders of magnitude [11] and leads to the emission of a broad band flash of light whose width is delineated in picoseconds.

Plasma processes have been invoked by a number of researchers as the basis for explaining the light emitting mechanism [5, 12-17]. Atoms that are neutral at low

2temperatures 'T' become ionized as T increases inside the collapsed bubble. In the resulting plasma charged particles emit and absorb light through various processes. Electrons moving with the thermal velocity can collide with neutral atoms or with ions to radiate light as they accelerate [thermal Bremsstrahlung]. Light propagating in the plasma can be absorbed by the inverse of these processes as well as by ionizing collisions with atoms. If the distance that light propagates from generation to re-absorption $l_\gamma$ is small compared to the size of the plasma then the medium is opaque and the spectrum will be that of a blackbody and radiation is from the surface [1]. For large values of $l_\gamma$ the body becomes transparent. Radiation is emitted form the volume of the bubble, so therefore the spectrum yields information about the underling collisional processes.

For hydrogen at 6000K and a density of 2.3x10$^{22}$ molecules/cm$^3$, $l_r \approx 10m$ as calculated from formulas for the inverse of electron-neutral Bremsstrahlung. This number is much greater than the .35$\mu m$ radius of the collapsed bubble [18] and therefore also greater than $R_e$. Similar calculations have led researchers to claim that the sonoluminescing medium is a transparent plasma [12-17,19]. Nevertheless, we observe the blackbody behavior shown in Figure 1. Reconciliation of theory and experiment must be sought in the fact that the contents of the SL bubble are very dense, being compressed to the van der Waal's hard core [5,11]. Saha's equation for the degree of ionization and the standard Bremsstrahlung formulas [including $l_r$] are calculated in the dilute gas limit where behavior is dominated by binary collisions. This could be the opposite limit to the one that describes SL [17]. A mechanism where a 'cool' blackbody spectrum can mask an arbitrarily high interior temperature [and shorter $l_r$] is pre-heating due to thermal radiation transport that is set off by a shock wave [1].

Another paradox of the blackbody interpretation of SL relates to the density of states. The derivation of Planck's formula assumes that many modes [i.e. many wavelengths of light] fit into the hot blackbody, but the SL hot spot is smaller than the wavelength of light that we measure. Perhaps those processes that lead to a small $l_r$ also lead to a large index of refraction [and therefore a smaller $\lambda$] inside the hot spot. In this model light would be released as $l_r$ suddenly increased in the cooling hot spot. The existence of such a switching process inside the bubble is suggested by the plot in Figure 2B which shows that the ratio of the ionization temperature to the blackbody temperature is the same value for all noble gas bubbles. That noble gas bubbles are accurately described by Planck's formula is shown in Figure 2A.

Measurements of the flash width as a function of color also support a switching mechanism. It is found that between 200nm and 700nm the flash width varies by less than 20% [being longer in the ultra-violet! {18}]. Within the absolute timing resolution of about 40ps it appears that the spectrum is blackbody from start to finish. Such behavior is also found to exist in plasmas that are generated at the focus of a strong laser in water [20].

A clue to the transition from opaque to transparent behavior is suggested by Figure 3 that displays the SL spectrum of a cloud of cavitating argon bubbles driven by sound at 1MHz [21]. The spectrum of light generated at this higher acoustic frequency is not well described by Planck's formula, but is characteristic of formulas for thermal

Bremsstrahlung[1]. The radii, $R_c$, of the 1MHz bubbles [21] at the moment of collapse are so small [$R_c \approx 50nm$] that perhaps these bubbles are transparent to light. Thus for SL the transition from opaque [blackbody] to transparent takes place at around 100nm. The robust behavior of SL even at the nanoscale has been pushed to 11MHz where scaling arguments suggest that $R_c \approx 10nm$. The spectral densities at 1MHz and 11MHz are similar [Figure 3].

The transduction of sound into light by a collapsing bubble is robust down to the nanoscale. Although radiation is emitted from a surface whose radius is smaller than, the wavelength of light and the photon matter mean free path, the observed spectrum can match that of a blackbody for [$200nm < \lambda < 800nm$ and $R_e > .1\mu m$]. Whether blackbody behavior extends out into the Rayleigh-Jeans limit of large $\lambda$ is a question for future experiments. As an equilibrated blackbody radiates only from its surface the extent to which the bubble's contents are stressed may be unrepresented in the observed spectrum. The spatial correlations and mode counting dilemma that would characterize such a small blackbody could constitute a new regime for the application of statistical optics. The inability to reconcile the long photon mean free path with the smallness of the hot spot suggests new physics in the modeling of SL.

8. All data in the paper is acquired from bubbles acoustically driven in sealed cylindrical resonators constructed with quartz walls and stainless steel end-caps [5]. Spectra were measured with light falling directly on the input slit (through order sorting filters) of a spectrometer (Acton 308i) read out by an intensified CCD (Princeton Inst. IMAX) and are fully radiance calibrated against commercially available QTH and $D_2$ lamps. Spectra in figures 1 and 3 acquired through commercial grade quartz (GM assoc. # 214) whereas fig2 is acquired in cells constructed of 'suprasil'. No spectra are corrected for transmission of water or quartz. For our commercial grade quartz, there is absorption for wavelengths below 300 nm that rises to 25 % at 200nm. The suprasil has constant transmission for all wavelengths above 200nm. We attribute the bump in the data at 550 nm and the dip at 360 nm to documented errors in the manufacturer-supplied calibration of our lamps (see fig 75 reference 5).

9. Spectra reported here have the same spectral density and detailed shape as reported in previous papers. But in the course of recalibrating the system we find that the scale for the 'y' axis, namely spectral radiance, is generally lower, being down by about a factor of 12 compared to 4,6. We have verified the new data calibrated against various lamp standards with photon counting through bandpass filters. Previously quoted values of photons per flash remain unchanged. We believe that the mistake in scaling the 'y' axis is greater than can be accounted for by resonator variability, drive level, and thermal drift [see discussion in ref.5]. The corrected value of radiance plus our ability to measure flash width and bubble size combine to make possible the quantitative comparisons to blackbody radiation proposed here.

22. Research supported by DARPA and the NSF.

FIGURE CAPTIONS

Figure 1: Spectrum of Sonoluminescence from a hydrogen bubble in water (23C) driven at 33 kHz. The hydrogen is dissolved into the water at a partial pressure of 5 torr. Data acquired with 24 nm FWHM resolution. The solid line is a fit to a blackbody at 6230 K. Using the measured flash width of 110 ps, this fit requires emission from a surface with radius of 0.22 microns. The dashed line is a bremsstrahlung fit with a temperature of 15000 K.

Figure 2: A) Spectrum of Sonoluminescence from bubbles of helium (150 torr) and xenon (3 torr) in water (23C) driven at 42 kHz. Resolution is 12 nm FWHM. The solid lines are blackbody fits at 8000K (xenon) and 20400K (helium). Using measured flash widths of 100 ps (helium) and 200 ps (xenon) gives emission from a surface of radius 0.1 microns (helium) and 0.4 microns (xenon). Ambient radii measured with light scattering techniques [5] are about 5.5 micron (xenon) and 4.5 micron (average value for helium) from which we estimate $R_c$ ($=R_0/7.6$ (xenon), $=R_0/9.8$ (helium)) as 0.7 micron (xenon) and 0.5 micron (helium). [Note that a 150Torr He bubble is not in diffusive equilibrium [5].] The dashed line is a bremsstrahlung spectrum at infinite temperature. B) Ratio of ionization potential ($\chi$) of the gas used to make sonoluminescence to the temperature of the blackbody ($T_{BB}$) best fitting the observed spectrum for each gas. This ratio is plotted vs. the ionization potential of the gas (k is Boltzman's constant). Gases plotted include the 5 light noble gases as well as hydrogen ($\chi=15.5$ eV for $H_2$).

Figure 3: Spectrum of Sonoluminescence from a cloud of cavitation bubbles in water driven at 1 MHz given per $cm^3$ of cloud volume. The water is saturated with argon and maintained at $18^0 C$. Resolution is 12 nm FWHM. The solid line is a bremsstrahlung fit with temperature 85,000 K and the dashed lines are blackbody curves with temperatures of 9900 K and 15,000K. The open circles are the measured spectrum (60 nm FWWM) of xenon bubbles driven in water (11C) with sound at 11 MHz and offset vertically by an arbitrary amount.

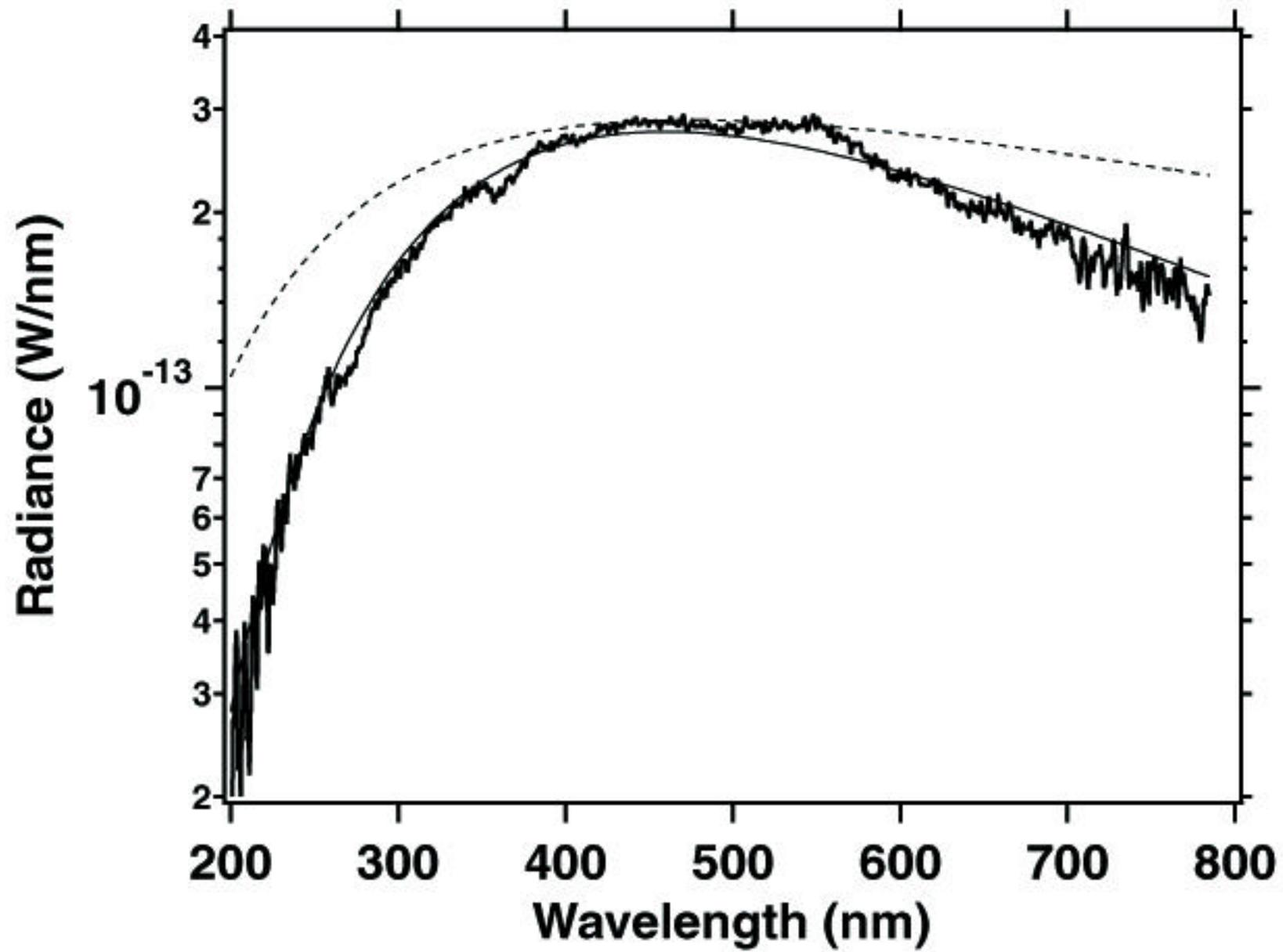

Figure 1 Vazquez

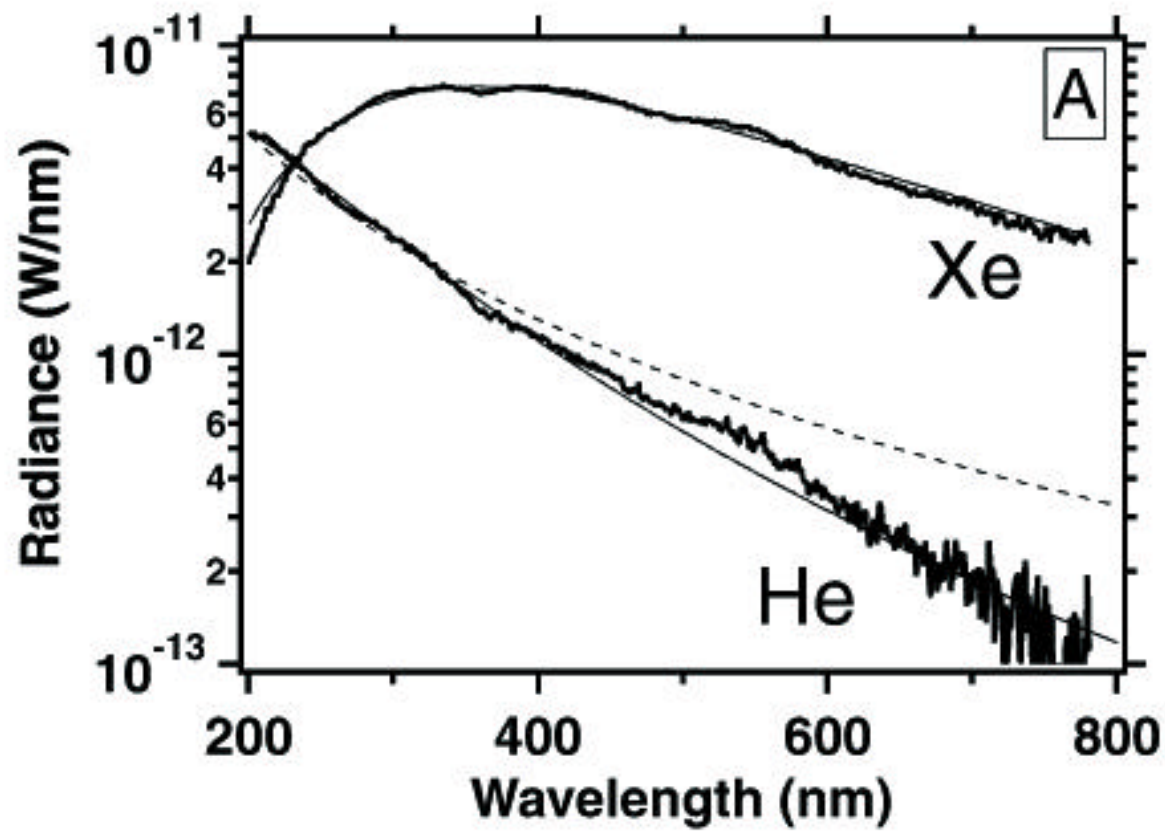
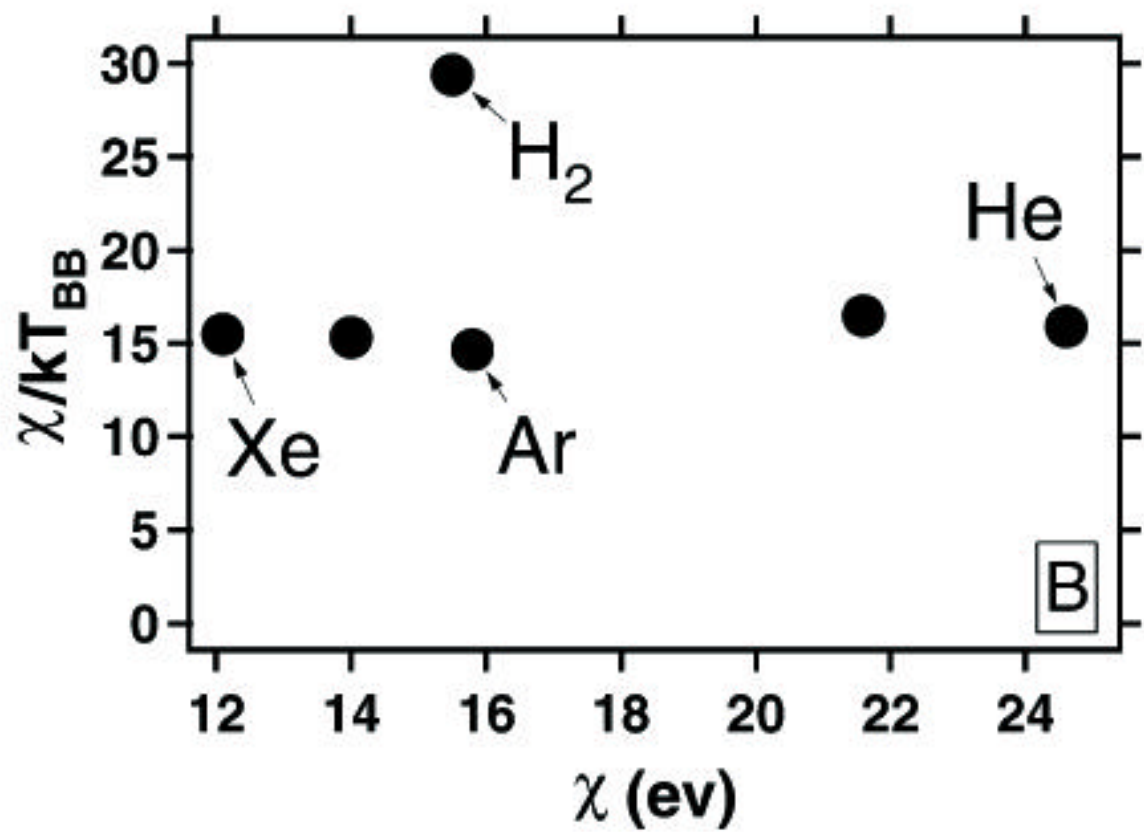

Figure 2 Vazquez

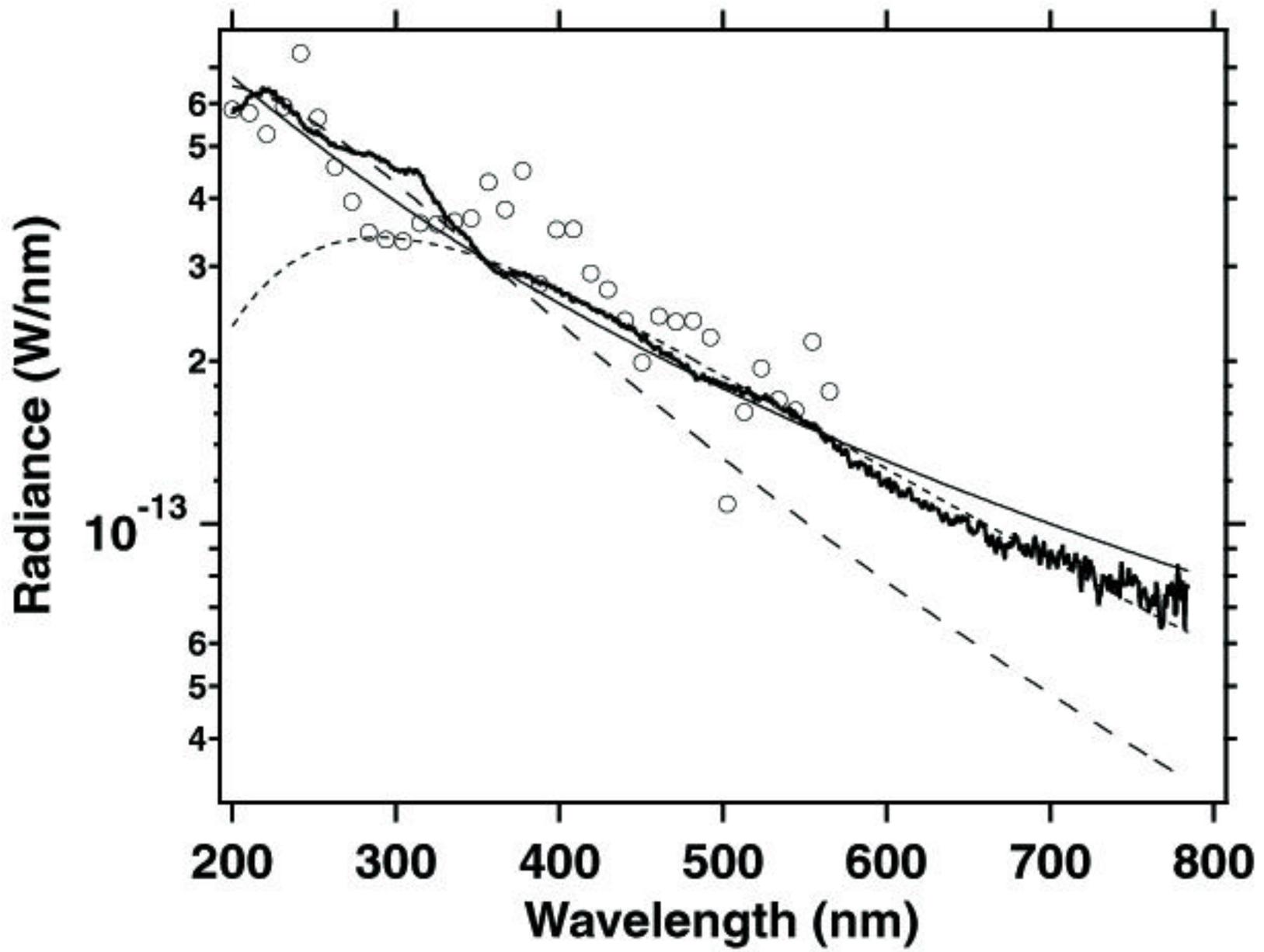

Figure 3 Vazquez